# Beyond headcount and human capital: The Effective Cognitive Population as a decomposable capacity unit for AI-era planning

Kwan Soo Shin [a,b]

[a] Department of AI and Innovation Management, Hanil University and Presbyterian Theological Seminary, Wanju, Republic of Korea

[b] PolymathMinds Lab, Asan, Republic of Korea

ORCID: 0009-0001-5799-7556

Correspondence: sshin@pmminds.ai

**Abstract**

National planning counts population, human capital, and artificial-intelligence preparedness in separate ledgers. Demographic accounting has advanced from headcount to skills-adjusted stocks and still debates how much age structure retains once skills are modeled, yet no existing unit carries the conditions under which preparedness becomes productive capacity. This study introduces the Effective Cognitive Population (ECP), a decomposable unit that weights population by capability and by the conditions under which capability is deployed, anchored to the World Bank Human Capital Index Plus (HCI+) and the non-overlapping dimensions of the IMF AI Preparedness Index. The architecture is portable in principle; the case tested here is artificial intelligence, which has a published preparedness index. For 144 countries, HCI+ becomes a productivity level, AI opportunity uses digital infrastructure and innovation integration, conversion governance uses regulation and ethics, and the benchmark is ECP = N H(1 + AC). Against 2024 total output on identical population bases, ECP raises criterion R-squared from 0.849 for the HCI+-adjusted stock to 0.882 and lowers leave-one-country-out RMSE from 0.723 to 0.641, with the working-age comparison identical and bootstrap intervals excluding zero. Eighty-nine of 144 countries move at least ten rank positions from headcount, mostly through the human-capital adjustment itself. Results are stable across denominators, vintages, aggregation forms, and a 27-rule multiverse. The direct A by C interaction is not statistically supported, so the conjunction is a planning rule rather than causal complementarity. ECP is a diagnostic ledger whose scope excludes forecasts of population decline and estimates of AI's causal productivity effect.

## 1. Introduction

Public planning depends on units of account. Population headcount determines the scale of service demand and the tax base; working-age population approximates potential labor supply; human-capital indices describe the productivity embodied in people; and technology-readiness indices describe the environment in which new tools may diffuse. Each unit is useful, but none by itself answers a planning question that has become central in the AI era: how much human productive capacity is available under a country's prevailing health, learning, employment, technological, and governance conditions?

The most obvious objections arrive before the construction does, and each is met by a design decision at the point of construction. Weighting population by capability is not new: human capital-weighted population estimates already cover 185 countries over a long horizon, so that weighting is inherited here rather than proposed [1]. Any composite of this kind then faces the standing charge that its weights encode trade-offs its users never see, which is answered by stating the trade-off in closed form, reporting the components beside the composite, and treating the weighting itself as an object of analysis [2]. And indices of artificial-intelligence preparedness track national income closely enough that their incremental content cannot be assumed [3], a gradient this study measures in its own sample, so the overlapping human-capital dimension is removed before combination, every criterion comparison conditions on capital intensity, and the resulting income gradient is stated as measured. Each response is tested at the point where it is used, and the tests are named in Section 3.

This is a measurement problem before it is a forecasting problem. A planning indicator must define the construct, prevent double counting, disclose its compensability assumptions, retain diagnostic decomposition, and remain stable under defensible alternative specifications. The composite-indicator literature has established that a high correlation with an outcome is not enough. Weighting, aggregation, coherence, uncertainty, and the distinction between statistical importance and normative weights must all be audited [4,5,6]. Recent work in *Socio-Economic Planning Sciences* has advanced data-driven weighting, multi-objective construction, benefit-of-the-doubt methods, and multiverse analysis for precisely these purposes [7,8,9,10], and has applied composite indicators directly to postshock industrial policy decisions [11].

The World Bank Human Capital Index Plus (HCI+) provides a strong starting point. HCI+ translates health, education, labor-market entry, employment, and learning at work into expected lifetime productivity under prevailing national conditions [12]. It improves on a simple schooling or test-score measure and has a direct log-productivity interpretation. It does not, however, represent the opportunity for AI-enabled augmentation or the governance conditions under which economy-wide adoption may occur. Conversely, the International Monetary Fund AI Preparedness Index (AIPI) measures digital infrastructure, innovation and economic integration, human capital and labor-market policy, and regulation and ethics [13]. Using the total AIPI together with HCI+ would count human capital twice. A defensible combined metric therefore requires decomposition rather than simple index addition.

There is also a theoretical reason not to treat AI readiness as an unconditional productivity bonus. The general-purpose-technology account holds that a technology of wide applicability raises output only where complementary innovation occurs downstream [14], and the measured version of that claim is that the complementary asset is organizational and largely unrecorded, so productivity appears to fall before it rises [15,16,17]. The task-based literature reaches the same conclusion from the labor side, since displacement and reinstatement of tasks proceed at rates that institutions and skills govern [18,19], and the surveys of AI and economic growth are explicit that the aggregate effect remains open rather than settled [20,21,22]. Measurement at intermediate scales agrees: exposure indices for occupations show wide variation in what AI can reach [23], firm-level instruments show that AI capability is an organizational construct rather than a technology stock [24], and field experiments find that the same tool produces different gains depending on how work is organized around it [25,26,27]. Management theory separately distinguishes automation from augmentation and explains why organizations face tensions between them [28]. At national scale, these mechanisms cannot be inferred from a single cross-section. A national metric can encode a conjunctive planning rule that discounts one-sided readiness, such as requiring both technological opportunity and conversion governance, but it must not present that rule as an estimated causal interaction.

One property of the resulting unit belongs in the open, because it is part of the claim. The architecture is not logically confined to artificial intelligence. Any technology whose returns depend on infrastructure and on a permissive regulatory regime could in principle be carried by a unit of the same form, wherever analogous opportunity and conversion measures exist [14], and that portability is the point of the design. The present study tests the architecture for artificial-intelligence preparedness alone, which is the case for which an international measure has been published. What makes the test consequential is itself measurable: among countries at the same level of human capital, digital infrastructure, innovation integration, and the regulatory conditions for economy-wide adoption differ widely, and Section 4 reports that dispersion directly.

This study develops the Effective Cognitive Population (ECP) around that distinction. ECP is a scenario-equivalent stock that adjusts either total or working-age population by (i) HCI+-implied relative lifetime productivity, (ii) AI opportunity measured without the AIPI human-capital component, and (iii) conversion governance measured from AIPI regulation and ethics. The measure is decomposable by construction. Its benchmark conjunctive rule penalizes one-sided readiness: high infrastructure and innovation do not receive the full AI augmentation when the governance term is weak. The rule is externally specified and bounded rather than estimated on gross domestic product (GDP).

The empirical contribution is a multi-part measurement audit on a common 144-country sample. First, same-base comparisons test whether the ECP extension adds concurrent criterion information beyond HCI+ when the population denominator is held fixed. Second, an incremental ladder and order-invariant grouped decomposition test what HCI+ and the AI opportunity-governance block add beyond physical capital intensity. Third, leave-one-country-out errors and paired bootstrap intervals assess whether apparent gains are driven by a small number of countries. Fourth, denominator, vintage, aggregation, amplitude, and exponent multiverses expose the sensitivity of ranks and criterion fit. Fifth, a residual screen demonstrates how the metric can identify countries for further planning diagnosis while explicitly testing, and failing to establish, simple external explanations for the residual.

The contribution is deliberately bounded, and it enters a live scholarly conversation rather than an empty field. This study does not infer the consequences of fertility decline from cross-country levels. Jones models the effect of a declining world research population on idea production [29]. Maestas, Mullen, and Powell estimate the effects of population aging on labor-force and productivity growth [30]. These are dynamic questions, whereas the estimand here is a contemporary national-capacity measure. Within demographic accounting, the conversation this study joins has moved steadily toward capability, and it remains open. Lutz and coauthors describe the age-structure account they set out to displace as one that "has become the dominant paradigm in the field of population and development" [31] (p. 12798), and report for 165 countries that "the results show a clear dominance of improving education over age structure and give evidence that the demographic dividend is driven by human capital" [31] (p. 12798), concluding that "investments in human capital bring the true demographic dividend" [31] (p. 12803). The same group later constructed a skills-adjusted measure of the adult stock on that premise [32]. Kotschy, Suarez Urtaza, and Sunde answered directly that "the demographic dividend is not a mere education dividend but the result of a complex interplay between shifts in the age structure and education composition" [33] (p. 25983). What both sides share is the assumption that the contest is between two demographic coordinates, age and schooling. Neither asks what technological conditions the resulting capability meets. Marois, Gietel-Basten, and Lutz showed with multidimensional projections that China's low fertility need not hinder future prosperity [34]. Most recently, Gu, Wu, Marois, Lutz, and Niu examined 336 Chinese cities over two decades and estimated, on a balanced panel of 289 of them, that task-based skill composition became the stronger of the two measured contributors to growth while remaining complementary to age-based labor supply [35]. This study does not retest those dynamic transitions, and a cross-section cannot settle a question about dynamics. The measurement carries one implication for that debate nonetheless: within quartiles of the working-age share, the capability measure still spans a ratio of up to 4.78 between its ninetieth and tenth percentiles, so an age quartile summarizes capability only coarsely. Both positions in that debate therefore require a capability coordinate measured independently of age structure, which is what the unit developed here supplies. It addresses an antecedent planning problem: constructing a transparent cross-national stock that uses HCI+-implied lifetime productivity as its human-capital base and keeps AI opportunity and a bounded governance condition jointly visible without double-counting human capital. The difference from the nearest existing units is therefore specific rather than rhetorical. Weighting population by human capability is itself established: skills-adjusted stocks and, most directly, human capital-weighted population estimates already convert headcount into a productivity-weighted quantity, the latter for 185 countries with long time coverage [1,32]. Those units are silent on the technological conditions in which the weighted capability is deployed. The distinction between holding a capability and being positioned to use it has a long lineage in growth accounting: Benhabib and Spiegel found that human capital enters cross-country outcomes substantially through the adoption and diffusion of technology rather than as an ordinary accumulable factor [36]. ECP carries that distinction into the unit itself by recording the conditions of deployment as separate entries. AI preparedness indices measure those conditions but are silent on the human stock that would use them, and they place human capital inside their own composite, so adding the two indices double counts it [13]. Construct-validation work on preparedness indices sharpens the point, since such indices track national wealth and governance closely enough that their incremental content cannot be assumed [3]. ECP therefore does not claim novelty for capability weighting. It contributes the joint account: it removes the overlapping dimension before combining, applies a conjunctive rule so that one-sided readiness is discounted, and reports the human, opportunity, and governance components separately, so a planner can read where a country's measured shortfall is concentrated rather than only its place in a single ranking. The long-established finding that country size has little systematic relation to per-capita outcomes [37,38] is treated only as a secondary check on returns to scale. Whether those channels operate remains an open empirical question, and posing it in panel form requires a contemporary capacity unit of the kind constructed here.

## 2. Measurement framework

### 2.1 Where the population unit stopped

Every national capacity measure answers one question before it answers any other: what counts as the productive population. The answer has been revised four times, and each revision fixed a defect in the previous unit while carrying an assumption forward. Setting out that sequence locates the specific place where the present study departs from it.

The first unit was the headcount. Population entered growth accounting as scale, and the strong version of the idea is that a larger population supports faster technological progress because ideas are nonrival, a mechanism formalized for the very long run by Kremer [39] and embedded in unified accounts of the transition out of Malthusian stagnation by Galor and Weil [40,41]. The unit is transparent and the mechanism is real over millennia. It stopped at the cross-section. Searches for a national scale effect in postwar data recovered no systematic relation between country size and per-capita outcomes [37,38], and the modern statement of the mechanism concerns the world stock of researchers rather than any single country's headcount [29].

Counting people therefore establishes the scale of demand and of the tax base while saying nothing about productive capacity.

The second unit conditioned on age. If dependents consume and workers produce, the ratio between them should govern what a population can generate, and the demographic-dividend literature built a substantial account on exactly that: transitional shifts in age structure explained a large share of the East Asian growth episode [42], and the framework was extended to ageing, saving, and the full national transfer accounts [43,44,45,46]. Its own contributors then supplied its correction. Sanderson and Scherbov showed that a fixed chronological threshold misstates ageing once remaining life expectancy is used instead [47,48], so the boundary between worker and dependent is itself a measurement choice rather than a fact. Age composition brackets capacity without determining it, which the present sample confirms directly in Section 4.4.

The third unit weighted people by human capital. Beginning from the theory of investment in people [49], it established that measured skill tracks national outcomes more closely than time enrolled [50,51,52] and supplied the country-level attainment datasets that made the weighting operational [53,54], while the returns literature established what an additional year of schooling is worth and how widely that value varies across individuals [55,56]. The World Bank Human Capital Index and its extension are the current institutional form of this stage, converting survival, schooling adjusted by learning outcomes, health, and employment into expected lifetime productivity [12,57]. The stage is well audited: comparisons built on it are sensitive to construction and uncertainty [58], and its subnational application shows how much variation a national figure conceals [59]. What it retains is the assumption that capability is a property of people considered apart from the tools they work with.

The fourth unit made the weighting multidimensional and forward-looking. Lutz and coauthors argued that improving education dominates age structure in carrying the dividend [31,60] and built population accounts stratified by age, sex, and education that made the claim testable [61,62,63], extended them to migration and productivity in Europe [64], to India [65], and to China [34,35], and reduced the apparatus to a single distributed quantity in the human capital-weighted population estimates for 185 countries [1]. That research program remains active and internally contested [66,67,68], and Kotschy and coauthors dispute its central claim [33]. Its instruments are the closest existing comparators to the measure developed here.

This is where the river stops. The fourth unit specifies with unusual precision who holds how much capability, and it can project that holding to 2100. It says nothing about the technological conditions under which the capability is deployed, which the unit leaves outside itself as a common background. The practical consequence is now visible: two countries with the same human capital-weighted population, facing the same AI frontier, differ in digital infrastructure, in innovation integration, and in whether a regulatory regime permits economy-wide adoption at all. Every unit above scores those two countries identically. The distinction between holding a capability and being positioned to use it has a long lineage in growth accounting, where human capital enters cross-country outcomes substantially through the adoption and diffusion of technology rather than as an ordinary accumulable factor [36,69,70,71]. The unit constructed below carries that distinction into the count itself.

### 2.2 From headcount to a scenario-equivalent capacity stock

Let $N_i$ denote the population of country i. Let $S_i$ denote its HCI+ score. HCI+ is reported on a scale equal to 100 times expected log lifetime productivity, with 325 representing the full-potential benchmark [12]. The score is therefore converted to a relative productivity level:

$$H_i = \exp\left(\frac{S_i - 325}{100}\right).$$

This transformation matters. Comparisons built on human-capital indices are themselves sensitive to construction and uncertainty [58], which is why the sensitivity programme below varies the HCI+ vintage, the aggregation form, and the weighting rule instead of holding the base index fixed. Multiplying population by the published HCI+ score would combine a count with a logarithmic index and would not yield a coherent stock. In contrast, $N_i H_i$ is interpretable as population scaled by lifetime-productivity potential relative to the HCI+ frontier. The same calculation with working-age population $N_i^{15-64}$ produces a workforce-based ledger.
The word cognitive in the construct's name refers to this human core, and specifically to its learning component: HCI+ adjusts expected years of schooling by harmonized learning outcomes, so the human weight reflects measured skill acquired rather than time enrolled [12]. The word carries no claim about individual cognition, and the AI terms introduced below describe conditions of deployment.

HCI+ remains a prospective measure. It describes the human capital that a child born under prevailing conditions can expect to accumulate, not the realized productivity of every current resident. Accordingly, $N_i H_i$ and ECP describe capacity under prevailing conditions rather than literal counts of current worker equivalents. They support relative comparison and policy-gap decomposition under prevailing conditions. They do not support arithmetic claims that a country possesses a specific number of fully productive persons.

### 2.3 AI opportunity and conversion governance without double counting

The AIPI consists of four additive dimension contributions: digital infrastructure (DI), innovation and economic integration (IEI), human capital and labor-market policies (HCLMP), and regulation and ethics (RE) [13]. The data service stores each contribution on an approximately 0 to 0.25 scale; their sum reproduces the published overall AIPI to numerical precision in the analysis panel. Each retained dimension is therefore multiplied by four to restore a 0 to 1 scale.

AI opportunity is defined as the equal-weight mean of the two opportunity dimensions:

$$A_i = \frac{4DI_i + 4IEI_i}{2} = 2(DI_i + IEI_i).$$

Conversion governance is defined as:

$$C_i = 4RE_i.$$

The HCLMP dimension is excluded because HCI+ already represents health, education, employment, and learning at work. This exclusion is not cosmetic. It prevents the combined measure from rewarding the same human-capital advantage once through HCI+ and again through the AIPI total.

The label conversion governance is intentionally narrower than conversion capacity. Regulation and ethics can support trusted adoption and diffusion, but they do not directly measure management quality, work redesign, decision rights, sectoral absorptive capacity, or worker-level complementarity. The observable $C_i$ therefore captures one auditable institutional margin of a broader organizational conversion capacity that remains only partly measurable.

### 2.4 Benchmark ECP and the amplitude grid

Per-capita capacity under the benchmark rule is:

$$Z_i(\lambda) = H_i(1 + \lambda A_i C_i),$$

and the corresponding ECP stock is:

$$ECP_i(\lambda) = N_i Z_i(\lambda).$$

The main specification sets lambda equal to 1. Since $A_i$ and $C_i$ lie between zero and one, the AI-governance augmentation factor then lies between one and two. This is a transparent normalization, not an estimate that frontier AI conditions double output. To expose the calibration choice, lambda values of 0, 0.25, 0.50, 0.75, and 1 are reported without selecting among them on GDP. Lambda equal to zero nests the HCI+-adjusted population comparator. The product $A_i C_i$ is a conjunctive rule. Recorded augmentation vanishes when either dimension is absent, and the marginal value of one dimension scales with the level of the other; in the aggregation taxonomy of composite indicators the rule penalizes imbalance multiplicatively without imposing the strict noncompensation of a minimum or an outranking rule, because partial compensation between nonzero levels remains. This rule is useful for planning because an arithmetic mean can allow exceptional infrastructure to compensate fully for weak governance, or the reverse. It is not equivalent to evidence that the structural cross-partial derivative of production with respect to A and C is positive. That empirical claim is examined separately and is not supported in the present cross-section.

A composite of this kind inherits a standing objection, and the design answers it in the terms the objection sets. Reviewing the mashup indices of development, Ravallion locates the defect precisely: "Little or no attention is given to whether the implied tradeoffs in the space of the primary dimensions being aggregated are defensible" [2] (p. 9). He states the consequence for exactly the decision problem addressed here, namely allocating public resources across the components of a composite: "Unless the mashup index considers, and reveals, its MRSs across components, or its marginal weights, it will be impossible to assess whether it is acceptable as a characterization of the development objective" [2] (p. 23). His preferred remedy is the "'dashboard' alternative of monitoring the components separately" [2] (p. 1).

The methodological literature supplies the instruments for meeting that demand. The handbook treatment sets out the sequence from normalization through weighting to aggregation and requires that uncertainty and sensitivity be propagated rather than reported after the fact [4,72,73], subsequent reviews map the space of weighting and aggregation choices and the interpretive commitments each carries [5,74], and critical assessments catalogue where the steps go wrong in practice [75] and how far rankings move when they are varied [76]. Within that space the present rule is a bounded-compensation form whose formal relatives are the generalized non-compensatory indices [77,78] and, on the optimization side, the benefit-of-the-doubt family and its hierarchical and cooperative-game extensions [79,80].

The construction answers on each count, and it adopts the remedy rather than arguing against it. The trade-off is stated in closed form, so the rate at which weak governance discounts strong opportunity is legible before any country is scored. The dashboard is retained, since the human, opportunity, and governance terms are reported beside the composite instead of disappearing into it. And the weighting is treated as an object of analysis through the amplitude grid and the 27-rule exponent multiverse of Section 3.4 rather than a single authored choice. What the composite adds to the dashboard is a common unit in which the three claims can be compared at all.

Several properties make the metric operationally useful. First, every term is bounded and decomposable. Second, no weight or component coefficient is estimated using the criterion outcome. Third, a planner can report ECP alongside H, A, and C, making a low score traceable to a specific margin. The same transparency also makes the metric falsifiable. If independent adoption and productivity data show that regulation and ethics do not function as a bottleneck, C must be replaced or the conjunctive rule abandoned. If the HCI+ productivity transformation ceases to transport across countries, the human-capital base must be recalibrated.

## 3. Data and methods

### 3.1 Data sources, timing, and sample

The analysis combines public international data at the country level. The intersection of the required variables contains 144 countries. Table 1 reports the role and reference period of each source.

**Table 1. Construct, operational measure, and data source**

| Construct | Operational measure | Reference period | Source and role |
|---|---|---|---|
| Total output Y | GDP, purchasing-power-parity, constant international prices | 2024 | World Development Indicators; concurrent criterion |
| Population N | Total population | 2024 | World Development Indicators; scale base |
| Working-age population N15-64 | Population aged 15 to 64 | 2024 | World Development Indicators; alternative scale base |
| Physical capital K | Real capital stock, variable rnna | 2023 | Penn World Table 11.0; capital-intensity control [81] |
| Human productivity H | HCI+ overall score transformed to a relative level | 2020 | World Bank HCI+; human-capital base [12] |
| AI opportunity A | Digital infrastructure plus innovation and economic integration | 2023 | IMF AIPI dimensions [13] |
| Conversion governance C | Regulation and ethics | 2023 | IMF AIPI dimension [13] |

World Development Indicators series are NY.GDP.MKTP.PP.KD, NY.GDP.PCAP.PP.KD, SP.POP.TOTL, and SP.POP.1564.TO. Total GDP was reconciled exactly to GDP per capita multiplied by population before estimation; the maximum absolute log identity gap after reconciliation is 1.78 times 10^-15. Capital is the Penn World Table 11.0 real capital-stock measure at constant national prices [81]. The 2020 HCI+ vintage is used in the main analysis so that it precedes the 2024 criterion year. The latest 2025 HCI+ vintage is reserved for sensitivity analysis. AIPI is a 2023 preparedness assessment released in 2024.

Inclusion follows from the component intersection alone; no outcome threshold enters the sample rule. No country is removed on the basis of its residual, rank, income group, or geopolitical status. All downloaded responses are retained verbatim, and the derived panel is hashed. Full country coverage and source hashes are reported in the Supplementary Material.

### 3.2 Concurrent criterion comparison

The first test asks whether ECP adds criterion information beyond the nearest comparator while holding the scale base constant. For each stock $U_i$, the model is:

$$\ln Y_i = b_0 + b_1 \ln U_i + e_i.$$

Seven units are compared on the identical 144-country sample: total headcount, HCI+-adjusted population, total-population ECP, working-age population, HCI+-adjusted workforce, working-age ECP, and an inherited formulation retained only as an audit benchmark. R-squared, adjusted R-squared, slope, and exact leave-one-country-out root-mean-square error (LOOCV RMSE) are reported. The comparison is concurrent criterion validity, not temporal forecasting.

For each pair of ECP and HCI+ units, the difference in LOOCV RMSE is resampled with a country-paired nonparametric bootstrap of 20,000 draws [82]. Pairing preserves the fact that both models are evaluated on the same held-out countries. Two-sided bootstrap probabilities smaller than the Monte Carlo resolution are reported as $p < 0.0001$.

### 3.3 Incremental and order-invariant validity

The second test conditions on physical capital intensity because development accounting already explains a large share of cross-country income variation through capital and human capital [57,83]. That tradition supplies the benchmark a new unit must clear: the aggregate production function with a residual [84], its augmented cross-country form carrying human capital as a third factor [85], and the accounting exercises that measure how much of the income gap the observed factors leave unexplained [86]. A capacity unit earns attention only when it carries information beyond those factors, which is what the ladder tests. Nested ordinary least-squares models use log GDP per capita as the criterion and HC1 heteroskedasticity-robust standard errors. The sequence adds HCI+, A, C, the A by C term, and population size to log capital per person. Model fit is assessed by R-squared, adjusted R-squared, Akaike information criterion, Bayesian information criterion, and exact LOOCV RMSE.

Sequential increments can depend on entry order. The HCI+ base and the three-term AI opportunity-governance block (A, C, and AC) are therefore also evaluated using an exact grouped Shapley decomposition over all block orders [87]. Contributions are calculated for R-squared and for reduction in cross-validated mean squared error. This decomposition assigns shared explanatory information without treating a normative composite weight as a regression importance weight [6].

The direct interaction test includes capital intensity, HCI+, A, C, AC, and population. Mean-centered and uncentered forms are compared, variance-inflation factors are reported, and the incremental interaction signal is checked with a 4,999-draw Freedman-Lane residual permutation. These tests distinguish the metric's externally imposed conjunction from an empirically estimated complementarity.

**3.4 Sensitivity and classification diagnostics**

Sensitivity analysis covers four sources of discretion. First, the scale base is changed from total to working-age population. Second, the 2025 HCI+ vintage replaces 2020. Third, the aggregation rule is changed to HCI+ only, ungated AI opportunity, square-root conjunction, additive A and C, and a broader governance measure that includes the AIPI human-capital and labor-policy dimension. The broad form is a sensitivity only because it reintroduces conceptual overlap. Fourth, the augmentation amplitude lambda varies from zero to one, and 27 theory-near exponent combinations vary the powers on H, A, and C over 0.75, 1, and 1.25. No rule is chosen by maximizing GDP fit.

Rank comparisons use Spearman correlation, absolute rank shifts, and the range of each country's rank across the 27-rule multiverse. To show that capacity is not reducible to age structure, countries are divided into quartiles of the working-age population share. The 10th, 50th, and 90th percentiles of per-capita capacity Z are reported within each quartile.

**3.5 The conversion residual and frontier sensitivity**

A descriptive conversion residual is obtained from:

$$\ln(Y_i/N_i) = c + \alpha\ln(K_i/N_i) + \gamma\ln Z_i + u_i.$$

Negative residuals identify cases where realized GDP per capita lies below the level associated with measured capital intensity and ECP capacity in this sample. The residual absorbs measurement error, omitted factors, temporary shocks, resource rents, and functional-form error. It is therefore a screening order for investigation, not an efficiency score or causal estimate.

As a model-form sensitivity, a half-normal stochastic frontier with normal noise is fitted by maximum likelihood [88]. Country inefficiency is calculated with the conditional estimator of Jondrow, Lovell, Materov, and Schmidt [89]. The likelihood-ratio statistic uses the boundary-mixture reference distribution, and ordering agreement with the OLS screen is measured by Spearman correlation. Natural-resource rents, government effectiveness, control of corruption, and regulatory quality are tested as external descriptive correlates. Failure to correlate is reported as a limitation rather than converted into a validation claim.

**3.6 Secondary check on returns to scale**

For continuity with development accounting, the composite is entered in:

$$\ln(Y_i/N_i) = c + \alpha\ln(K_i/N_i) + \gamma\ln Z_i + \delta\ln N_i + \varepsilon_i.$$

The population coefficient is tested for equivalence within a prespecified negligible elasticity interval of -0.05 to 0.05 using a 90% confidence interval [90]. Algebraically, the per-capita equation can be rewritten as $\ln Y_i = c + \alpha \ln K_i + \gamma \ln Z_i + (1 - \alpha + \delta)\ln N_i$. Thus delta equal to zero is a constant-returns accounting condition given Z. This is a secondary diagnostic. It neither identifies the dynamic mechanisms studied by Jones [29] and Maestas, Mullen, and Powell [30] nor overturns the scale evidence in Backus, Kehoe, and Kehoe [37] and Rose [38].

## 4. Results

### 4.1 Distribution and internal coherence

The 144-country sample spans the full observed development range. HCI+ scores range from 81.66 to 283.50. After the productivity-level transformation, H ranges from 0.088 to 0.660. AI opportunity ranges from 0.204 to 0.778, and conversion governance from 0.121 to 0.922. Benchmark per-capita capacity Z ranges from 0.091 to 1.105. A value above one is possible because H is expressed relative to the HCI+ full-potential benchmark while the AI term represents an additional scenario layer.

The four IMF dimension contributions sum to the published overall AIPI with a maximum absolute discrepancy of 1.31 times 10^-14. This identity check confirms that the rescaling of the retained AIPI dimensions recovers their intended 0 to 1 dimension scales. Removing HCLMP from the AI block also makes the conceptual accounting explicit: HCI+ supplies human productivity; A supplies technological opportunity; and C supplies the narrower governance condition.

### 4.2 Like-for-like criterion comparison

Table 2 and Figure 1 compare each capacity unit against 2024 total output. Headcount alone accounts for 62.3% of the cross-country variation and has a LOOCV RMSE of 1.142 log points. Multiplying the same population base by HCI+-implied productivity raises R-squared to 0.849 and lowers LOOCV RMSE to 0.723. Adding the benchmark AI opportunity-governance term raises R-squared to 0.882 and lowers LOOCV RMSE to 0.641.

The comparison is not driven by using total population for ECP and working-age population for the comparator. On the working-age base, HCI+-adjusted workforce produces R-squared of 0.873 and LOOCV RMSE of 0.664, while working-age ECP produces 0.902 and 0.582. The ECP slopes are close to one under both bases.

**Table 2. Like-for-like concurrent criterion validity for alternative national-capacity units**

| Unit | R-squared | Adjusted R-squared | LOOCV RMSE | Slope |
|---|---|---|---|---|
| Headcount N | 0.623 | 0.620 | 1.142 | 0.855 |
| HCI+-adjusted population N H | 0.849 | 0.848 | 0.723 | 1.002 |
| ECP population N H(1 + AC) | 0.882 | 0.881 | 0.641 | 1.001 |
| Working-age population N15-64 | 0.656 | 0.654 | 1.091 | 0.886 |
| HCI+-adjusted workforce N15-64 H | 0.873 | 0.872 | 0.664 | 1.016 |
| ECP working age N15-64 H(1 + AC) | 0.902 | 0.902 | 0.582 | 1.011 |
| Inherited ECP specification | 0.795 | 0.794 | 0.842 | 0.967 |

*Note.* N = 144 for every row. The criterion is log 2024 PPP total GDP. LOOCV RMSE is exact leave-one-country-out root-mean-square error in log output. The inherited specification is included as an audit benchmark and is not used elsewhere.

The paired bootstrap difference in LOOCV RMSE between total-population ECP and HCI+-adjusted population is -0.0828, with a 95% interval of -0.0971 to -0.0677 ($p < 0.0001$). The working-age difference is -0.0824, with a 95% interval of -0.0974 to -0.0668 ($p < 0.0001$). These are internal cross-country criterion gains. They do not demonstrate future forecasting accuracy.

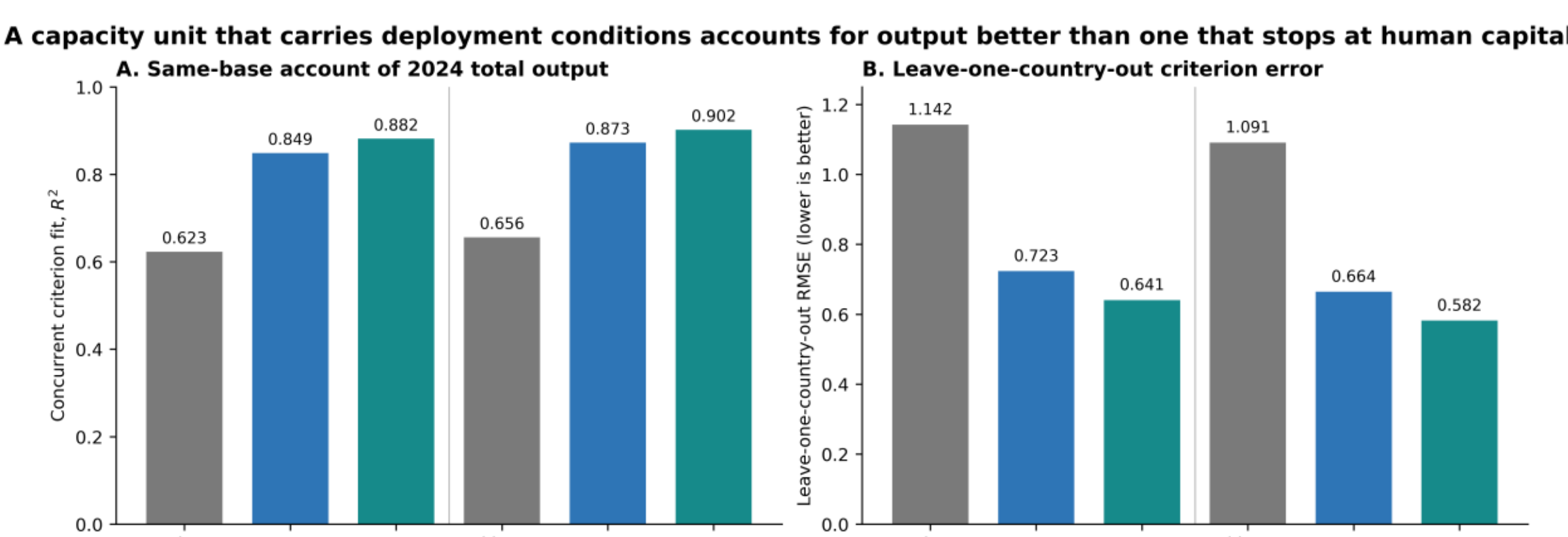

Figure 1. A capacity unit that carries deployment conditions accounts for output better than one that stops at human capital. The comparison is between units: each bar is the same population base counted a different way, so the difference isolates what the unit records. Panel A reports R-squared from log 2024 PPP total GDP regressed on the log of each capacity unit. Panel B reports exact leave-one-country-out RMSE; lower values are better. Every comparison uses the same 144 countries. Both panels are drawn on a zero baseline and every value is printed above its bar. Gray denotes unadjusted population, blue denotes HCI+-adjusted population, and teal denotes benchmark ECP.

### 4.3 What the components add beyond capital intensity

The conditional results are more demanding. Capital intensity alone yields R-squared of 0.9065 and LOOCV RMSE of 0.3585 for log GDP per capita (Table 3). HCI+ lowers the error to 0.3207. AI opportunity lowers it to 0.3168, and conversion governance lowers it further to 0.3137. Adding the direct AC interaction raises the error to 0.3160 even though in-sample R-squared rises slightly. Population then raises it again to 0.3176.

**Table 3. Incremental validity conditional on capital intensity**

| Model | R-squared | Adjusted R-squared | AIC | BIC | LOOCV RMSE | Change in RMSE |
|---|---|---|---|---|---|---|
| M0: capital intensity | 0.9065 | 0.9058 | -295.5 | -289.6 | 0.3585 | |

| Model | R-squared | Adjusted R-squared | AIC | BIC | LOOCV RMSE | Change in RMSE |
|---|---|---|---|---|---|---|
| M1: + HCI+ productivity | 0.9265 | 0.9254 | -328.1 | -319.2 | 0.3207 | -0.0378 |
| M2: + AI opportunity A | 0.9292 | 0.9277 | -331.6 | -319.7 | 0.3168 | -0.0039 |
| M3: + conversion governance C | 0.9317 | 0.9298 | -334.8 | -320.0 | 0.3137 | -0.0031 |
| M4: + direct interaction AC | 0.9323 | 0.9298 | -334.0 | -316.2 | 0.3160 | +0.0023 |
| M5 scale check: + population | 0.9328 | 0.9299 | -333.1 | -312.3 | 0.3176 | +0.0016 |

*Note.* N = 144. The criterion is log GDP per capita. Change in RMSE is relative to the preceding row. Coefficient inference uses HC1 standard errors.

The order-invariant decomposition reaches the same substantive conclusion. Conditional on capital intensity, the HCI+ block contributes 0.0113 to R-squared and 0.0143 to cross-validated mean-squared-error reduction. The AI opportunity-governance block contributes 0.0145 and 0.0144, respectively. The latter block contains A, C, and AC jointly, so this contribution cannot be read as evidence for the interaction itself.

The direct interaction coefficient is -1.075 (HC1 t = -0.90, p = 0.371), and the Freedman-Lane permutation probability is 0.378. Centering leaves the coefficient and inference unchanged; the centered interaction VIF is 1.25. Accordingly, the study does not claim that national-level complementarity has been detected. The benchmark conjunction remains a conjunctive measurement decision. Relative to the more permissive stock N H(1 + A), it lowers LOOCV RMSE by 0.0129, with a paired-bootstrap 95% interval of -0.0197 to -0.0065, but this comparison validates a constrained rule against the same criterion; identification of a production mechanism lies beyond it.

The capital control does not depend on which Penn World Table series measures capital. The reported specification uses the stock at constant national prices, the series constructed for comparison within a country over time; the stock at current purchasing-power parities is the series constructed for comparison across countries. Re-estimating with the latter matches all 144 countries and leaves the two log capital-intensity measures correlated at 0.999998, so the variance and error statistics of Table 3 are unchanged at the precision they are reported to, moving by at most 0.0003 in R-squared and 0.0005 in leave-one-country-out RMSE. The information criteria shift by at most 0.41 on a scale where adjacent models differ by tens of points, so the ordering of the ladder is identical, and the conjunctive coefficient becomes -1.032 with a t-ratio of -0.88 (Table S12).

The benchmark unit also imposes a common elasticity on its three factors, and that restriction is directly testable. Regressing log output on log population, log human-capital weight, and the log augmentation factor without restriction gives elasticities of 0.946, 1.562, and 1.598, and joint equality is rejected, with a chi-squared statistic of 87.3 on two degrees of freedom. Decomposing the rejection locates it. Equality between the human-capital weight and the augmentation weight is not rejected, with a chi-squared statistic of 0.002 on one degree of freedom and a p-value of 0.97, whereas equality between the population elasticity and the human-capital weight is rejected, with a statistic of 12.97 and a p-value of 0.0003. The proportionality that ECP newly imposes, between capability weighting and AI augmentation, is therefore the one component of the common-elasticity restriction that this test does not reject. Failing to reject an equality is weaker than evidence for it, and the result is reported at that strength. What the test does reject is the common exponent linking scale to quality, which Section 4.6 examines as an accounting question and Section 4.4 reports across the exponent multiverse. A freely estimated model is unavailable as a planning unit in any case, because its weights are taken from the criterion the unit is later used to interpret and would move with sample and year. The comparison is reported to locate what the fixed unit assumes (Table S13).

### 4.4 Rank reclassification, age structure, and specification uncertainty

Headcount and ECP remain related because population is one factor of ECP, but their rank correlation is 0.902 rather than one. Eighty-nine of 144 countries move by at least ten positions; the median absolute movement is 13 and the maximum is 42. Attribution across stages matters: seventy-six of those eighty-nine already move at least ten positions under the HCI+ adjustment alone, and the median additional movement contributed by the AI extension over the HCI+-adjusted ranking is two positions. The reclassification finding therefore belongs to capability adjustment as a whole, while the AI extension's distinctive contributions are the criterion-fit gain and the component-level diagnostics. Singapore, Switzerland, Denmark, Finland, New Zealand, Sweden, Norway, and Ireland record the largest upward movements. Chad, Guinea, Mali, Burundi, Sierra Leone, the Republic of Congo, Malawi, and Senegal record the largest downward movements. These changes should be read as differences between a scale ledger and a capacity-adjusted ledger; welfare comparison lies outside their scope.

Capacity dispersion remains wide within comparable age structures. Within quartiles of the working-age population share, the ratio of the 90th to the 10th percentile of Z is 1.83, 4.78, 4.45, and 3.17 from the lowest to the highest quartile. Age composition and capacity are therefore non-substitutable planning dimensions. Demographic structure informs labor-force and productivity-growth analysis [30], while education- and skills-adjusted measures show why the capability of a population cannot be inferred from age composition alone [31,32].

Because AI preparedness indices track national income closely [3], the reclassification is also examined by income group. The augmentation factor is strongly graded by income: group means rise from 0.095 in low-income countries through 0.135 and 0.220 to 0.454 in high-income countries, and the Spearman correlation between income group and the augmentation factor is 0.849 across the 143 countries carrying an income classification. The same table qualifies what that gradient implies. Within every income group the augmentation factor still spans a ratio between 2.37 and 3.22 from its tenth to its ninetieth percentile, so income brackets the factor while leaving substantial variation inside each bracket. And the reranking that the AI block produces relative to the HCI+ stock is largest at the bottom of the income distribution: the median absolute movement is 3.5 positions among low-income countries, where 37.5 percent of countries move at least five positions, against 1.0 position and 5.1 percent among upper-middle-income countries and 3.0 positions and 34.6 percent among high-income countries, with a Kruskal-Wallis statistic of 22.1 across the four groups and a p-value of 0.0001. The income gradient is therefore substantial and is reported as measured. Two further features of the same table are descriptive rather than interpretive: variation inside each band remains wide, and the reranking is not monotone in income (Table S14).

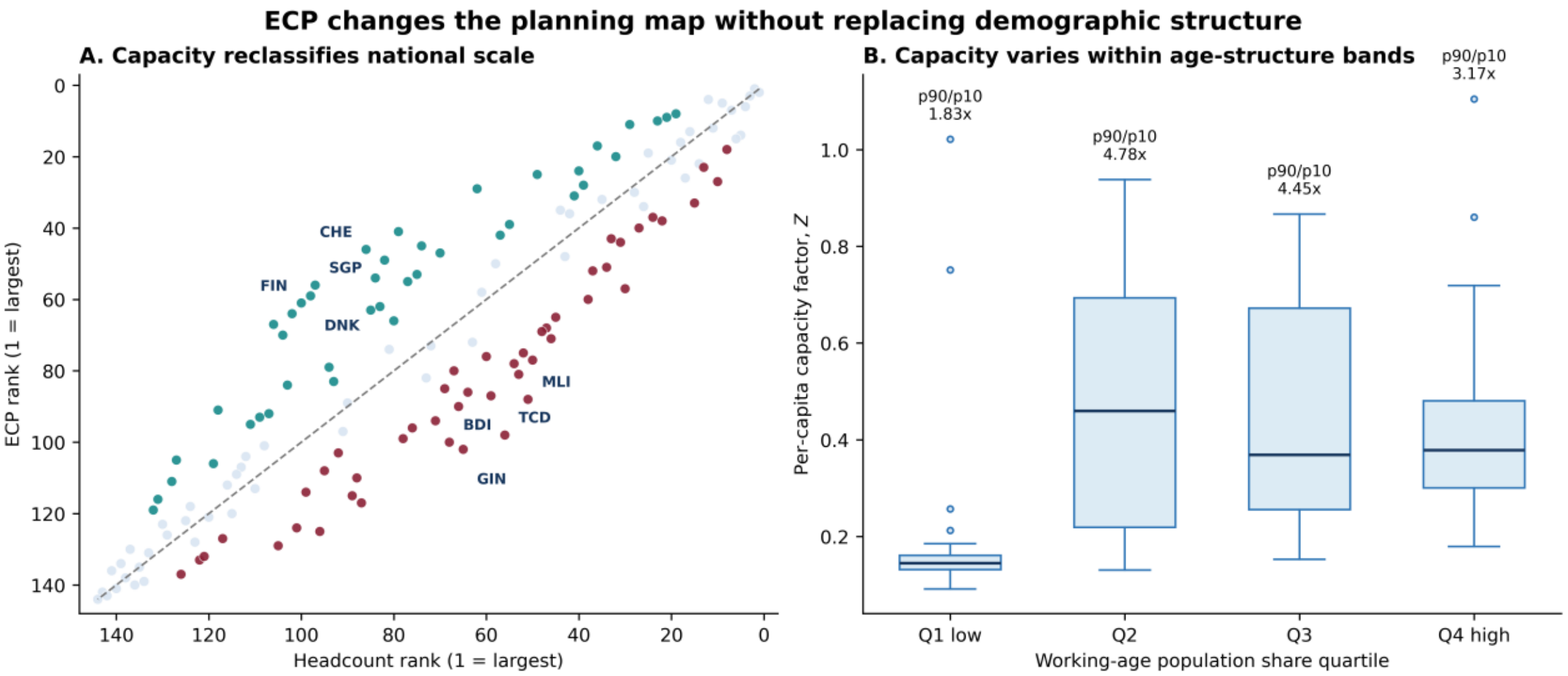

Figure 2. ECP changes the planning map without replacing demographic structure. Panel A compares descending headcount and ECP ranks. Teal points move upward under ECP and crimson points move downward; selected large shifts are labelled. Panel B reports benchmark

per-capita capacity Z across quartiles of the working-age population share. The annotated ratios compare the 90th and 10th percentiles within each quartile.

The conclusions survive the declared specification changes. Using working-age population yields a rank correlation of 0.998 with the total-population benchmark and a median absolute rank shift of one. Alternative aggregation rules yield correlations of 0.999 or higher with the benchmark except the HCI+-only form, which still yields 0.996. Across the 27 exponent combinations, the median country rank range is six positions, the 90th percentile is 12, and the maximum is 19.

The augmentation amplitude analysis clarifies what is and is not identified. Relative to HCI+ alone at lambda equal to zero, even lambda equal to 0.25 lowers LOOCV RMSE from 0.723 to 0.697; successive fixed scenarios of 0.50, 0.75, and 1 produce errors of 0.675, 0.656, and 0.641. These scenarios are not outcome-selected, and the monotonic pattern is not a license to choose a larger lambda. It shows that the criterion result is not dependent on granting the full benchmark augmentation. The 2025 HCI+ sensitivity, on 141 countries, similarly lowers LOOCV RMSE from 0.714 for the HCI+ stock to 0.628 for ECP.

### 4.5 A bounded residual for conversion gaps

The residual regression has R-squared of 0.929 and LOOCV RMSE of 0.314. The most negative residuals occur for Haiti, Ukraine, Tajikistan, the Central African Republic, Sudan, the Kyrgyz Republic, Jordan, Latvia, Madagascar, and Lebanon. The most positive occur for Gabon, Egypt, Guinea, Angola, Luxembourg, Saudi Arabia, Pakistan, Zimbabwe, Eswatini, and Mali. This ordering is plausible enough to motivate diagnosis but not clean enough to validate a single mechanism. Conflict, institutional disruption, rents, informal production, price measurement, and omitted industrial structure can all enter the residual.

The stochastic-frontier sensitivity converges with a maximum absolute numerical gradient of $1.05 \times 10^{-5}$. It estimates $\sigma_u = 0.368$, $\sigma_v = 0.209$, and $\gamma = 0.755$. The likelihood-ratio statistic against symmetric OLS noise is 6.377, with a boundary-mixture p value of 0.0058. More importantly for the limited purpose of this analysis, the SFA inefficiency order correlates 0.995 with the OLS under-conversion order. The screen is therefore not an artifact of using a symmetric residual, although neither model establishes inefficiency without stronger environmental controls.

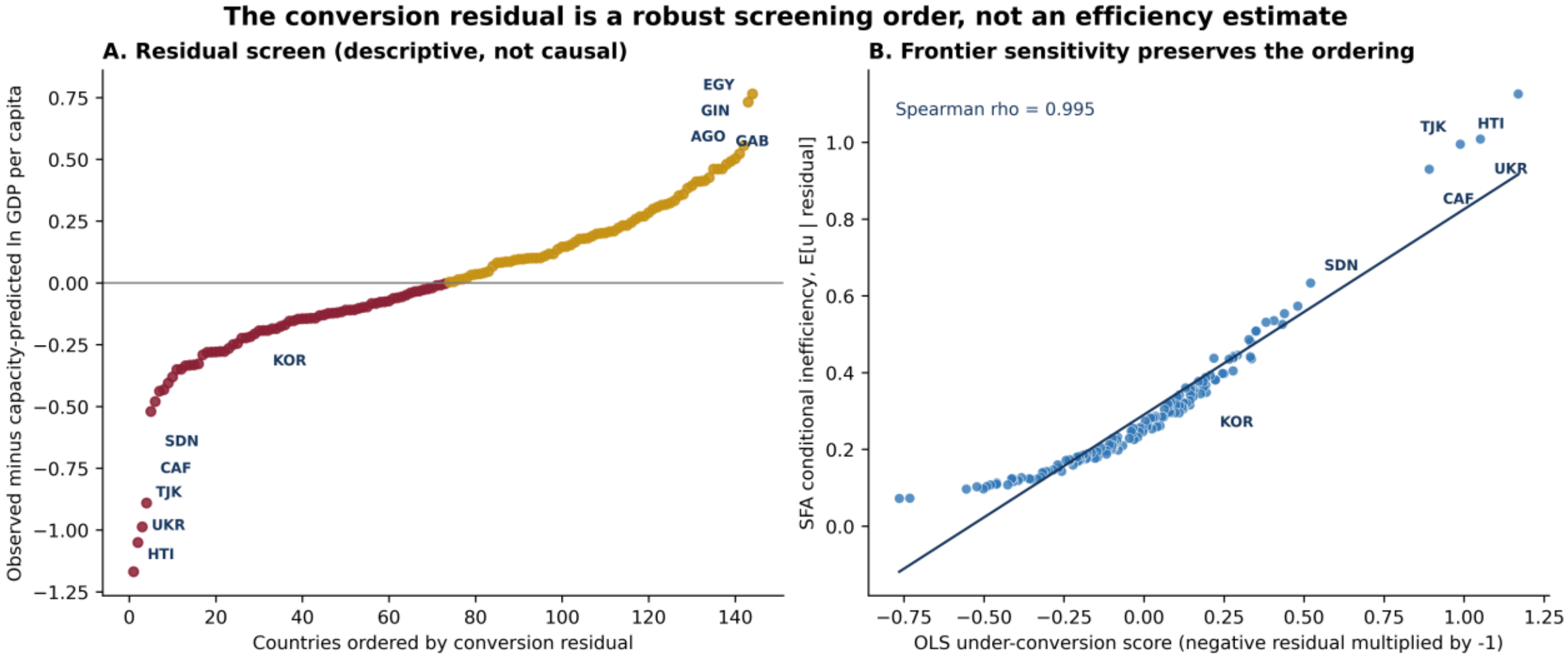

Figure 3. The conversion residual provides a robust screening order whose interpretation stops short of an efficiency estimate. Panel A orders countries by the residual from log GDP per capita on capital intensity and log Z. Panel B compares the OLS under-conversion score with conditional inefficiency from a half-normal stochastic frontier. The close ordering agreement does not establish causal inefficiency.

External diagnostics reinforce that caution. The conversion residual is not significantly associated with natural-resource rents (rho = 0.013, p = 0.874), government effectiveness (rho = 0.101, p = 0.231), control of corruption (rho = 0.111, p = 0.185), or regulatory quality (rho = 0.112, p = 0.180). Korea's residual is -0.194, the 30th most negative of 144 under OLS and 39th under SFA. Korea therefore illustrates where the screen points investigators; its position is neither extreme in the distribution nor a validated mechanism. Firm- and sector-level evidence would be required to determine whether organizational conversion, industrial composition, or measurement accounts for the gap.

### 4.6 The scale check is not a demographic result

In the secondary composite specification, the coefficient on log Z is 0.541 ($t = 6.11$), and the residual population coefficient is -0.011 ($t = -0.86$). Its 90% confidence interval, -0.032 to 0.010, falls inside the prespecified equivalence bounds of -0.05 and 0.05. Model R-squared is 0.930 and LOOCV RMSE is 0.315.

This result supplies positive evidence for a negligible residual scale term within the accounting model. It does not show what happens when a country's population falls. Backus, Kehoe, and Kehoe [37] and Rose [38] already documented the weakness of national scale effects. Jones's [29] idea-based growth mechanism operates through the world research population, and Maestas, Mullen, and Powell [30] estimate age-structure dynamics rather than contemporary national size. The present estimate is compatible with these literatures and does not overturn them.

## 5. Discussion

### 5.1 What ECP contributes

ECP contributes a planning architecture rather than a new production law. Its human-capital base inherits the productivity interpretation and life-course coverage of HCI+. Its AI extension separates technological opportunity from conversion governance and excludes the AIPI human-capital component to avoid overlap. Its stock can be calculated on total or working-age population, and every result can be decomposed back to H, A, and C.

The strongest evidence is comparative and measurement-specific. On the same countries and the same population base, the ECP extension adds concurrent criterion information beyond HCI+. The improvement is visible both in fit and in held-out-country error, persists under a conservative augmentation amplitude, and appears on total-population and working-age ledgers. Conditional on capital intensity, HCI+, AI opportunity, and conversion governance each reduce LOOCV error. These results meet a more demanding standard than showing that a large composite correlates with GDP, but they remain internal validation against a criterion to which several inputs are conceptually related.

One comparison in the supplement clarifies what the unit assumes rather than what it fits, and it belongs in the main argument. A freely estimated three-regressor model outperforms every fixed unit in Table 2. That is expected, because its weights are taken from the criterion the unit is later used to interpret and would move with sample and year, which is what disqualifies it as a planning unit. The informative part is where the fixed restriction breaks. Equality between the human-capital weight and the augmentation weight, which is the proportionality ECP newly imposes, is the one component that test does not reject, and failing to reject it is weaker than evidence for it. The rejection falls instead on the common exponent linking scale to quality, a feature this unit inherits from every capability-weighted stock that preceded it (Table S13).

The metric changes planning diagnosis in two ways. First, it separates national scale from the productive potential associated with prevailing human and AI conditions. The resulting rank shifts show which countries look larger or smaller when capacity is counted, without implying that the ECP rank is a welfare league table. Second, the component ledger locates the margin on which a shortfall is recorded. H is linked to health, learning, and employment over the life course; A to digital infrastructure and innovation integration; C to regulation and ethics. A government can therefore distinguish a human-capital constraint from an opportunity constraint or a governance constraint.

The same transformation serves decisions taken outside national government, and it does so without further assumptions, because each use reads the same decomposition rather than a new estimate. Multilateral lenders and aid agencies allocate across countries on population or income denominators, and a capacity-adjusted denominator changes which countries appear underserved relative to what their populations could produce. Subnational planners face the identical accounting problem one level down, and the transformation applies wherever the component indicators exist at that level, which subnational human-capital measurement has already shown to be feasible [59]. Firms selecting locations for skill-intensive operations screen on workforce size and wage levels, and the component ledger separates a country whose people are capable but whose deployment conditions are weak from one where both are weak, which are different investment cases with different time horizons.

**5.2 The conjunctive rule and its evidentiary boundary**

The conjunctive term is both the most distinctive and the most easily overstated part of the measure. In composite-indicator design, aggregation schemes that limit compensation are justified when a deficit in one essential dimension should not be erased by strength in another [4,5]. Munda and Nardo set out the defensible setting for that choice, arguing that linear aggregation assumes full substitutability among dimensions and that a country ranking should avoid imposing substitutability where it does not hold [77]. The rule adopted here sits inside that setting without occupying its strict end: it discounts one-sided readiness continuously and retains partial compensation between nonzero levels. That is the motivation used here; a minimum-type or outranking gate would impose strict noncompensation, and inverted benefit-of-the-doubt formulations offer a further route to limited compensation within this journal's methodological toolkit [91]. Comparing those forms against the rule used here is a separate measurement question with its own criterion. It is compatible with research on complementary intangible investment [15], heterogeneous human-AI performance [25,26], and the automation-augmentation tension [28], but consistency is not identification. The distinction between measured preparedness and realized performance recurs well beyond artificial intelligence: a reassessment of the Global Health Security Index found that higher scored preparedness accompanied higher per-capita mortality early in the COVID-19 pandemic, contrary to expectation, with the association reversing as the pandemic progressed, and concluded that the validity of a forward-looking preparedness score depends on what governments subsequently do as well as on the capability being scored [92]. That is the general reason a capacity ledger should keep opportunity and the conditions for its conversion visible as separate entries.

The direct interaction test is null and slightly negative. It would therefore be incorrect to write that this cross-section verifies an A by C productivity mechanism. The benchmark rule instead asks a policy question: how much AI-related capacity should be credited when both opportunity and governance are present? The answer is reported as a bounded scenario family, and the raw components remain visible. Independent longitudinal adoption, organizational change, and productivity data are required to estimate the interaction. Persistent absence of complementarity in those data would refute the gate and require an additive or alternative bottleneck form.

### 5.3 Relation to human-capital and demographic measurement

HCI+ supplies the productivity scale on which ECP rests, and ECP adds to that scale what capability-weighted stocks currently leave outside the unit: the technological and governance conditions under which the weighted capability is deployed. HCI+ already combines health, formal learning, tertiary completion, employment, and learning at work [12]. Skills-adjusted and prospective-age measures likewise show that demographic capacity cannot be read from age or headcount alone [31,32,47], and human capital-weighted population estimates now supply such a weighted stock for 185 countries over a long horizon [1]. ECP inherits that weighting rather than proposing it. The added value of ECP is specific: it makes AI opportunity and one governance condition visible in the same national ledger while preserving their separation from human capital.

This positioning also resolves the mismatch that arises when a cross-sectional level measure is used to make a dynamic population claim. Population dynamics may affect world idea production [29], age composition, labor supply, and productivity growth [30]. Education and skills can alter the capacity associated with a given demographic structure [31,32]. ECP does not remove those channels. It offers a measure with which future panel or scenario research can ask whether changes in human capability and AI conversion moderate them. Until such designs exist, ECP should not be used to claim that fertility decline is harmless or that cognitive capacity substitutes for population at an identified rate.

### 5.4 Limitations and research program

Five limitations set the boundary of the present evidence. First, the study is cross-sectional. LOOCV holds out countries, not future years, and no coefficient identifies a temporal or causal effect. Second, HCI+ is prospective. Multiplying it by current population creates a scenario-equivalent stock under prevailing conditions, an accounting construct that stands apart from any observed current labor input. Cohort-specific human-capital stocks would be preferable for dynamic work.

Third, criterion overlap and reverse causation remain. Richer countries can invest more in human capital, infrastructure, innovation, and regulation, while HCI+ is constructed to carry a productivity interpretation. Construct-validation work on national AI readiness indices makes this concern concrete, reporting that such indices covary closely with wealth and governance quality so that their incremental content beyond income cannot be assumed [3]. The design responds on each point, and every response is testable: the human-capital dimension of the preparedness index is removed before combination, the opportunity and governance components are reported separately instead of summed into a single readiness score, and every criterion comparison conditions on capital intensity. Section 4.4 reports the corresponding evidence. The augmentation factor is graded by income, as that literature anticipates, and it retains a spread of between 2.37 and 3.22 inside every income group while reranking low-income countries more than high-income ones. Usage-based evidence reaches the same territory from another direction: constructing measures of observed AI use across countries from five waves of a usage index, Fan and Nguyen value the labor time currently saved by AI at 2.7 trillion dollars a year and report that both income and regulatory readiness predict how concentrated that value is within a country [93]. Measured preparedness and realized use are therefore related without being interchangeable, which is the condition under which a separate governance entry earns its place. The residual risk that the retained components still track development level is one reason the direct interaction test is reported as null rather than as support. GDP criterion performance is therefore evidence of coherence and incremental information, not proof of a distinct causal factor. Validation against independent outcomes, such as sectoral AI diffusion, task-level productivity, patent quality, or firm-level organizational redesign, is required.

Fourth, conversion governance is incomplete. Regulation and ethics may enable trusted diffusion, but they do not directly measure management practice, workflow redesign, implementation skill, or the distribution of AI use across firms. The weak external correlations of the residual confirm that the present screen has not isolated a conversion mechanism. A dedicated, independently measured conversion-capacity instrument is the highest-priority extension.

Fifth, the aggregation rule is normative. The lambda grid and 27-rule multiverse quantify sensitivity, but they do not reveal a true universal weight. Different planning tasks may warrant different calibration. The recommended practice is to publish the component dashboard and scenario range together with the benchmark ECP, never the rank alone.

These limitations define a research sequence rather than a retreat from the construct. The next test should join time-varying AI adoption and productivity to independent measures of management and organizational redesign at country-industry or firm level. A second extension should build cohort-specific ECP stocks and test within-country long differences. A third should examine whether ECP-informed resource allocation improves policy decisions relative to headcount, working-age population, HCI+, or AIPI alone. Each test can falsify a different part of the framework.

## 6. Conclusion

National planning needs a ledger that distinguishes how many people a country has from the productive potential created by prevailing human-capital and AI conditions. ECP supplies that ledger by converting HCI+ to its proper productivity level, excluding duplicate human-capital information from AIPI, and adding a transparent AI opportunity-governance scenario term. Across 144 countries, the extension adds concurrent criterion information beyond HCI+ on identical population bases, materially changes the map of national scale, and remains stable across a wide specification audit.

The result is neither a causal estimate of AI productivity nor a verdict on population decline. The direct interaction is not supported, and the conversion residual is a screen rather than an efficiency measure. Those negative findings sharpen the contribution: ECP is a decomposable, auditable planning construct whose assumptions are visible and whose mechanism can be tested with future longitudinal and micro-level evidence.

## Declarations

### Author contribution

K.S.S. conceived the framework, designed the study, curated the data, developed the methodology and software, performed the analysis, prepared the visualizations, and wrote the manuscript.

### Funding

This research received no specific grant from any funding agency in the public, commercial, or not-for-profit sectors.

### Competing interests

The author declares no competing interests.

### Data and code availability

All source data are public and are identified in Table 1 and the Supplementary Material. A reproducibility archive containing the constructed analysis panel, raw-source request records where redistribution is permitted, executable Python scripts, environment requirements, source hashes, a data dictionary, a lineage record, and a one-command regeneration script is archived at Zenodo (https://doi.org/10.5281/zenodo.21871061; the version archived at submission is 1.0.0,

https://doi.org/10.5281/zenodo.21871062). The runner checks its inputs before it starts and its outputs after it finishes, and it fails rather than reporting success if any required file is absent or short. Every reported number is regenerated by that one command and written to a machine-readable result ledger, with the single exception of the capital-measure robustness table, which takes one further command because the Penn World Table file is distributed by its owners. Generative-AI assistance (Anthropic Claude) was used in the research process for code development, computational execution, and statistical checking under the author's specification, and every reported number was verified by the author against the ledger, so no result depends on unverifiable AI output. The archive includes superseded artifacts and the record of corrections.

**Ethics statement**

The study uses aggregate country-level public data and involves no human participants, identifiable personal information, animals, or biological materials. Institutional ethics approval and informed consent were not applicable.

**Declaration of generative AI and AI-assisted technologies in the writing process**

During the preparation of this work, the author used Anthropic Claude to assist with manuscript refinement, computational experiments, and statistical analysis. The author reviewed and edited the resulting material and takes full responsibility for the content of the publication.

# Supplementary Material

**Beyond headcount and human capital: The Effective Cognitive Population as a decomposable capacity unit for AI-era planning**

Kwan Soo Shin

Department of AI and Innovation Management, Hanil University and Presbyterian Theological Seminary, Wanju, Republic of Korea

PolymathMinds Lab, Asan, Republic of Korea

Correspondence: sshin@pmminds.ai

## S1. Purpose and evidentiary boundary

This supplement documents the complete construction and validation of the submission-version Effective Cognitive Population (ECP). It is limited to a 144-country cross-section. All reported holdout errors are leave-one-country-out errors within that cross-section. They are not time-series forecasts. No analysis in the manuscript or supplement identifies the effect of fertility, population decline, AI adoption, regulation, or human-capital investment.

The final measure differs materially from the inherited exploratory formulation. It uses the World Bank Human Capital Index Plus (HCI+) as the human-productivity base, converts the published log-productivity score to a level before constructing a stock, decomposes the International Monetary Fund AI Preparedness Index (AIPI), and excludes AIPI's human-capital and labor-market-policy dimension from the AI term. The inherited formulation is retained in one comparison row solely as an audit trail. It is not used to support the manuscript's conclusions.

## S2. Reproducible source alignment

### S2.1 Source series

| Variable | Series or field | Main reference year | Construction |
|---|---|---|---|

| Variable | Series or field | Main reference year | Construction |
|---|---|---|---|
| GDP | WDI NY.GDP.MKTP.PP.KD | 2024 | PPP total output, reconciled to GDP per capita times population |
| GDP per capita | WDI NY.GDP.PCAP.PP.KD | 2024 | PPP output per person |
| Total population | WDI SP.POP.TOTL | 2024 | Count |
| Population aged 15 to 64 | WDI SP.POP.1564.TO | 2024 | Count |
| Real capital stock | PWT 11.0 rnna | 2023 | Divided by total population for capital intensity |
| HCI+ | WB_HCIP, HD_HCIP_OVRL | 2020 | Main human-productivity score |
| HCI+ latest vintage | WB_HCIP, HD_HCIP_OVRL | 2025 | Sensitivity only |
| Overall AIPI | IMF DataMapper AI_PI | 2023 | Identity check only |
| Digital infrastructure | IMF DataMapper DI | 2023 | Contribution multiplied by four |
| Innovation and economic integration | IMF DataMapper IEI | 2023 | Contribution multiplied by four |
| Human capital and labor-market policy | IMF DataMapper HCLMP | 2023 | Excluded from main ECP; broad-governance sensitivity only |
| Regulation and ethics | IMF DataMapper RE | 2023 | Contribution multiplied by four |

The IMF component contributions reproduce the overall AIPI by summation. In the final panel, the maximum absolute discrepancy between AI_PI and DI + IEI + HCLMP + RE is $1.31 \times 10^{-14}$ and the mean absolute discrepancy is $4.00 \times 10^{-15}$.

### S2.2 Timing rule

The criterion year is 2024. HCI+ 2020 is the main human-capital vintage because it precedes the criterion year; HCI+ 2025 is a sensitivity. AIPI has one published reference year, 2023. Capital is the latest PWT observation preceding the criterion year, 2023. This alignment is not a causal lag design. It is a transparent ordering that avoids using the 2025 HCI+ score as the main input for 2024 output.

### S2.3 Identity reconciliation

The downloaded WDI GDP and GDP-per-capita series can differ slightly because of source updates and aggregation conventions. For the final panel, total output is reconciled as GDP per capita multiplied by population. After reconciliation:

- maximum absolute difference between ln(GDP/population) and ln(GDP per capita): $1.78 \times 10^{-15}$;
- mean absolute difference: $2.38 \times 10^{-17}$.

This prevents a mechanical denominator inconsistency in the total-output and per-capita specifications.

### S2.4 Frozen hashes

| Artifact | SHA-256 |
|---|---|
| Original inherited country panel | c89d09a56b79020e0d021b6ddcfef403ad1f00b43435bd616bf5b1e6e026dd91 |
| 2024 aligned 149-country panel | 0c8784f0829a71ecf066265d1b12631b44b891d9ea9cf8c946587616dcdf707b |
| Final 144-country analysis panel | 9d331314862196efc0ecd00b8f5c34575252fc0c186137438bcc8c70fa7d436e |
| IMF AI_PI raw response | 5a2aadb6596e3cbfe198e1b86032b3218d31117ea4b375acd676d9116a634bf5 |
| IMF DI raw response | 83a66ee3c3a7c69db6a0775c94243e93ac5981c55689d0e6c8da54ab63faacee |
| IMF IEI raw response | b42180421da1d4945a65d63545b5d0f4526ea6697ff0f089f2f17e14545583d4 |
| IMF HCLMP raw response | 48440fac8abde385152a3a5b1a20fcf9f048958ed95cac2503e21914a74f6c4f |
| IMF RE raw response | 56e209e3a392cd99fbed37e1b9f45eecad508672c705845aafc2fa023d4887a4 |
| World Bank HCI+ raw response | 57c24afab314ad025872ffc1faea429d86ece6bbeb1eea057021886812a80be0 |

Hashes for all distributed files are regenerated in the submission manifest.

## S3. Measurement construction

### S3.1 Human-productivity level

Let S denote the HCI+ score. The World Bank publishes HCI+ as 100 times an expected log lifetime-productivity quantity and sets the full-potential score to 325. The comparable level is therefore:

$$H = \exp[(S - 325)/100].$$

This gives H = 1 at the HCI+ full-potential benchmark. The observed H range is 0.0877 to 0.6603. Because HCI+ is prospective under prevailing conditions, H is a scenario-level productivity factor rather than an observation of current average worker productivity.

**S3.2 Non-overlapping AIPI dimensions**

The IMF API stores each of the four AIPI dimension contributions on an approximately 0 to 0.25 scale. The main ECP constructs:

$$A = \text{mean}(4DI, 4IEI) = 2(DI + IEI),$$

$$C = 4RE.$$

Observed A ranges from 0.2041 to 0.7783 and C from 0.1206 to 0.9218. HCLMP is not used because its health, skills, and labor-policy content overlaps conceptually with HCI+.

**S3.3 Benchmark and alternatives**

The benchmark is:

$$Z = H(1 + AC), \qquad ECP = NZ.$$

The working-age version replaces N with N15-64. The amplitude family is:

$$Z(\lambda) = H(1 + \lambda AC), \quad \lambda \in \{0, 0.25, 0.50, 0.75, 1\}.$$

Lambda is not estimated on GDP. Lambda equal to one is the benchmark normalization; lambda equal to zero is HCI+ alone. The alternative-form audit uses:

1. HCI+ only: NH;
2. ungated opportunity: NH(1+A);
3. square-root conjunction: NH(1+sqrt(AC));
4. additive dimensions: NH[1+0.5(A+C)];
5. broad governance: $NH[1 + AC_{broad}]$, where $C_{broad} = 2(HCLMP + RE)$;

6. working-age base: N15-64 H(1+AC).

The broad form is deliberately labelled a sensitivity because it reintroduces overlap with HCI+.

### S3.4 Descriptive distribution

Displayed values are rounded once, half-up, from the full-precision statistic.

| Variable | Mean | SD | Minimum | 25th percentile | Median | 75th percentile | Maximum |
|---|---|---|---|---|---|---|---|
| HCI+ score, 2020 | 187.873 | 53.683 | 81.660 | 139.800 | 193.732 | 233.256 | 283.498 |
| H productivity level | 0.291 | 0.148 | 0.088 | 0.157 | 0.269 | 0.400 | 0.660 |
| AI opportunity A | 0.471 | 0.154 | 0.204 | 0.341 | 0.450 | 0.607 | 0.778 |
| Conversion governance C | 0.515 | 0.191 | 0.121 | 0.367 | 0.516 | 0.626 | 0.922 |
| Per-capita capacity Z | 0.393 | 0.252 | 0.091 | 0.176 | 0.337 | 0.523 | 1.105 |

## S4. Statistical procedures

### S4.1 Exact leave-one-country-out error

For ordinary least squares, the held-out residual for country i is computed exactly as $e_i/(1 - h_{ii})$, where $e_i$ is the full-sample residual and $h_{ii}$ is leverage. LOOCV RMSE is the square root of the mean squared held-out residual. This avoids refitting 144 models and is algebraically identical to explicit leave-one-out refitting for linear regression.

### S4.2 Paired bootstrap

Each model supplies 144 squared held-out residuals. Country indices are sampled with replacement 20,000 times using seed 20260805. In each draw, RMSE is calculated for both models on the identical resampled country indices and differenced. Percentile intervals use the 2.5th and 97.5th percentiles. The two-sided probability is twice the smaller empirical tail probability, capped at one.

### S4.3 Robust inference and permutation

Coefficient standard errors use HC1. The direct AC interaction is also evaluated with a Freedman-Lane test. The reduced model includes capital intensity, HCI+, A, C, and population. Its residuals are permuted 4,999 times, added back to the reduced fitted values, and the full interaction t statistic is recalculated. The p value includes the observed statistic through the standard plus-one correction.

### S4.4 Grouped Shapley decomposition

Capital intensity remains in every model. Two blocks are permuted over all possible orders: (i) HCI+ and (ii) the AI opportunity-governance system A, C, and AC. For each order and performance functional, a block receives its marginal contribution when entered. Contributions are averaged across orders. The functionals are R-squared and negative LOOCV mean squared error, so a positive value indicates improved performance.

### S4.5 Stochastic frontier

The frontier is ln GDP per capita = X beta - u + v, where X contains an intercept, log capital per person, and log Z; u is half-normal and nonnegative; and v is normal. Maximum likelihood begins from five initial gamma values, uses Powell search, and is polished with L-BFGS-B. Convergence requires optimizer success, finite likelihood, and maximum absolute numerical gradient below 0.001. Conditional E[u|epsilon] follows Jondrow et al.

## S5. Complete statistical results

### Table S1. Paired-bootstrap differences in held-out-country RMSE

| Comparison | Difference | 95% interval | Two-sided p |
|---|---|---|---|
| Total-population ECP minus HCI+-adjusted population | -0.0828 | [-0.0971, -0.0677] | <0.0001 |
| Working-age ECP minus HCI+-adjusted workforce | -0.0824 | [-0.0974, -0.0668] | <0.0001 |
| Conjunctive ECP minus ungated opportunity stock | -0.0129 | [-0.0197, -0.0065] | <0.0001 |

A negative value favors the first unit named.

### Table S2. Incremental-model coefficients with HC1 standard errors

| Model | Predictor | Coefficient | HC1 SE | p |
|---|---|---|---|---|
| M0 | ln capital per person | 0.7909 | 0.0215 | <0.0001 |
| M1 | ln capital per person | 0.5830 | 0.0395 | <0.0001 |
| M1 | HCI+ / 100 | 0.6211 | 0.1092 | <0.0001 |
| M2 | ln capital per person | 0.5743 | 0.0397 | <0.0001 |
| M2 | HCI+ / 100 | 0.3827 | 0.1389 | 0.0066 |

| Model | Predictor | Coefficient | HC1 SE | p |
|---|---|---|---|---|
| M2 | AI opportunity A | 0.9831 | 0.4038 | 0.0162 |
| M3 | ln capital per person | 0.5772 | 0.0389 | <0.0001 |
| M3 | HCI+ / 100 | 0.3441 | 0.1304 | 0.0092 |
| M3 | AI opportunity A | 0.3672 | 0.4310 | 0.3958 |
| M3 | conversion governance C | 0.6556 | 0.3188 | 0.0416 |
| M4 | ln capital per person | 0.5675 | 0.0432 | <0.0001 |
| M4 | HCI+ / 100 | 0.3280 | 0.1267 | 0.0107 |
| M4 | AI opportunity A | 1.0618 | 0.8713 | 0.2251 |
| M4 | conversion governance C | 1.1626 | 0.7422 | 0.1195 |
| M4 | AC | -1.0342 | 1.1754 | 0.3805 |
| M5 | ln capital per person | 0.5589 | 0.0448 | <0.0001 |
| M5 | HCI+ / 100 | 0.2927 | 0.1426 | 0.0421 |
| M5 | AI opportunity A | 1.3255 | 0.9658 | 0.1722 |
| M5 | conversion governance C | 1.1208 | 0.7585 | 0.1418 |
| M5 | AC | -1.0749 | 1.1970 | 0.3708 |
| M5 | ln population | -0.0176 | 0.0176 | 0.3191 |

Intercepts are included in estimation and archived in final_incremental_coefficients.csv but omitted from the display. M5 is the additive-component bridge; the composite scale check is reported separately in Table S8.

### Table S3. Order-invariant grouped contributions beyond capital intensity

| Block | Contribution to R-squared | Contribution to LOOCV MSE reduction |
|---|---|---|
| HCI+ human-capital base | 0.0113 | 0.0143 |
| AI opportunity-governance system | 0.0145 | 0.0144 |
| Total | 0.0258 | 0.0287 |

### Table S4. Aggregation and denominator sensitivity

| Form | Rank correlation with benchmark | Median absolute rank shift | R-squared | LOOCV RMSE |
|---|---|---|---|---|
| Benchmark conjunction | 1.0000 | 0.0 | 0.8816 | 0.6406 |
| HCI+ only | 0.9960 | 2.0 | 0.8489 | 0.7234 |
| Ungated AI opportunity | 0.9992 | 1.0 | 0.8768 | 0.6535 |
| Square-root conjunction | 0.9996 | 0.0 | 0.8785 | 0.6490 |
| Additive A and C | 0.9995 | 0.0 | 0.8784 | 0.6492 |
| Broad governance | 0.9997 | 0.0 | 0.8798 | 0.6453 |
| Working-age denominator | 0.9976 | 1.0 | 0.9023 | 0.5819 |

R-squared and LOOCV RMSE use log total GDP as the concurrent criterion. The working-age row changes the population base and should be compared with the corresponding working-age HCI+ comparator in the main text.

### Table S5. AI-augmentation amplitude scenarios

| Lambda | Rank correlation with lambda = 1 | Median absolute rank shift | R-squared | LOOCV RMSE | Slope |
|---|---|---|---|---|---|
| 0.00 | 0.9960 | 2.0 | 0.8489 | 0.7234 | 1.0020 |

| Lambda | Rank correlation with lambda = 1 | Median absolute rank shift | R-squared | LOOCV RMSE | Slope |
|---|---|---|---|---|---|
| 0.25 | 0.9976 | 1.0 | 0.8597 | 0.6970 | 1.0028 |
| 0.50 | 0.9986 | 1.0 | 0.8685 | 0.6750 | 1.0026 |
| 0.75 | 0.9995 | 1.0 | 0.8756 | 0.6565 | 1.0018 |
| 1.00 | 1.0000 | 0.0 | 0.8816 | 0.6406 | 1.0007 |

The amplitude grid was specified as a reporting grid and is not optimized on GDP. Values greater than one are not searched because the purpose is to test whether the benchmark result survives conservative attenuation, not to maximize criterion fit.

### Table S6. Largest rank movements from headcount to ECP

| Direction | Country | Headcount rank | ECP rank | Change |
|---|---|---|---|---|
| Gain | Singapore | 97 | 56 | +41 |
| Gain | Switzerland | 86 | 46 | +40 |
| Gain | Denmark | 98 | 59 | +39 |
| Gain | Finland | 100 | 61 | +39 |
| Gain | New Zealand | 106 | 67 | +39 |
| Gain | Sweden | 79 | 41 | +38 |
| Gain | Norway | 102 | 64 | +38 |
| Gain | Ireland | 104 | 70 | +34 |
| Loss | Chad | 56 | 98 | -42 |
| Loss | Guinea | 65 | 102 | -37 |
| Loss | Mali | 51 | 88 | -37 |
| Loss | Burundi | 68 | 100 | -32 |
| Loss | Sierra Leone | 87 | 117 | -30 |
| Loss | Congo, Rep. | 96 | 125 | -29 |
| Loss | Malawi | 53 | 81 | -28 |
| Loss | Senegal | 59 | 87 | -28 |

Ranks are descending, with one denoting the largest stock. Across all countries, Spearman rho is 0.9025, the median absolute shift is 13, 89 countries move at least ten positions, and the maximum movement is 42.

### Table S7. Per-capita capacity within working-age-share quartiles

| Quartile | N | 10th percentile | Median | 90th percentile | 90th/10th ratio |
|---|---|---|---|---|---|
| Q1, lowest working-age share | 36 | 0.109 | 0.145 | 0.199 | 1.83 |
| Q2 | 36 | 0.184 | 0.460 | 0.877 | 4.78 |
| Q3 | 36 | 0.186 | 0.369 | 0.825 | 4.45 |
| Q4, highest working-age share | 36 | 0.217 | 0.378 | 0.687 | 3.17 |

### Table S8. Composite scale check

| Quantity | Estimate |
|---|---|
| N | 144 |
| R-squared | 0.9296 |
| LOOCV RMSE | 0.3154 |
| Coefficient on ln Z | 0.5411 |

| Quantity | Estimate |
|---|---|
| HC1 t on ln Z | 6.11 |
| Residual population coefficient delta | -0.0111 |
| HC1 t on delta | -0.86 |
| 90% confidence interval for delta | [-0.0322, 0.0101] |
| Equivalent within [-0.05, 0.05] | Yes |

The equivalence result describes the residual scale term in this accounting specification. It is not a within-country population effect.

### Table S9. Conversion-screen extremes

| Most negative | Residual | Most positive | Residual |
|---|---|---|---|
| Haiti | -1.169 | Gabon | 0.765 |
| Ukraine | -1.051 | Egypt, Arab Rep. | 0.732 |
| Tajikistan | -0.988 | Guinea | 0.555 |
| Central African Republic | -0.891 | Angola | 0.522 |
| Sudan | -0.521 | Luxembourg | 0.502 |
| Kyrgyz Republic | -0.480 | Saudi Arabia | 0.492 |
| Jordan | -0.438 | Pakistan | 0.480 |
| Latvia | -0.432 | Zimbabwe | 0.461 |
| Madagascar | -0.406 | Eswatini | 0.460 |
| Lebanon | -0.381 | Mali | 0.460 |

### Table S10. External correlates of the conversion residual

| Variable | Reference year | N | Spearman rho | p |
|---|---|---|---|---|
| Natural-resource rents, percent of GDP | 2021 | 143 | 0.013 | 0.874 |
| Government effectiveness | 2022 | 144 | 0.101 | 0.231 |
| Control of corruption | 2022 | 144 | 0.111 | 0.185 |
| Regulatory quality | 2022 | 144 | 0.112 | 0.180 |

These weak results prevent the conversion residual from being interpreted as a validated institutional-efficiency measure.

### Table S11. Stochastic-frontier sensitivity

| Diagnostic | Result |
|---|---|
| Converged | Yes |
| Maximum absolute numerical gradient | 1.053 times 10^-5 |
| Sigma u | 0.3680 |
| Sigma v | 0.2094 |
| Lambda, the ratio of Sigma u to Sigma v | 1.757 |
| Gamma | 0.755 |
| LR statistic versus normal OLS | 6.377 |
| Boundary-mixture p | 0.0058 |
| Spearman correlation, SFA versus OLS under-conversion order | 0.9949 |
| Korea SFA rank from most inefficient | 39 of 144 |
| Korea OLS rank from most negative | 30 of 144 |

**Table S12. Capital-measure robustness: constant national prices versus current PPPs**

Penn World Table 11.0 supplies two capital-stock series. The series at constant national prices is built for comparison within a country over time; the series at current purchasing-power parities is built for comparison across countries. All 144 sample countries match to both for 2023, and the two log capital-intensity measures correlate at 0.999998. The ladder of Table 3 is re-estimated with each.

| Model | R-squared constant national prices | R-squared current PPPs | LOOCV constant national prices | LOOCV current PPPs |
|---|---|---|---|---|
| M0: capital intensity | 0.9065 | 0.9062 | 0.3585 | 0.3590 |
| M1: + HCI+ productivity | 0.9265 | 0.9264 | 0.3207 | 0.3210 |
| M2: + AI opportunity A | 0.9292 | 0.9291 | 0.3168 | 0.3170 |
| M3: + conversion governance C | 0.9317 | 0.9316 | 0.3137 | 0.3139 |
| M4: + conjunctive term A*C | 0.9323 | 0.9322 | 0.3160 | 0.3162 |
| M5 bridge: + population size | 0.9328 | 0.9327 | 0.3176 | 0.3177 |

The direct conjunctive coefficient in M4 is -1.034 with a t-ratio of -0.880 under constant national prices and -1.032 with a t-ratio of -0.877 under current PPPs. No inference in the manuscript depends on the choice. As a provenance check, the maximum absolute difference between the analysis panel's capital column and the Penn World Table constant-national-prices series for 2023 is exactly zero.

**Table S13. Testing the unit's common-elasticity restriction**

The benchmark unit multiplies population, the human-capital weight, and the augmentation factor, which imposes a common elasticity on all three when log output is regressed on the log composite. The unrestricted model relaxes that constraint. A freely estimated model is not an alternative planning unit, because its weights come from the criterion the unit is later used to interpret; the comparison identifies what the fixed unit assumes.

| Specification | R-squared | Adjusted R-squared | LOOCV RMSE |
|---|---|---|---|
| Restricted: ln ECP | 0.8816 | 0.8807 | 0.6406 |

| Specification | R-squared | Adjusted R-squared | LOOCV RMSE |
|---|---|---|---|
| Free: ln N, ln H, ln(1+AC) | 0.9281 | 0.9266 | 0.5070 |

| Free elasticity | Estimate | HC1 standard error |
|---|---|---|
| ln N | 0.9458 | 0.0247 |
| ln H | 1.5619 | 0.1782 |
| ln(1+AC) | 1.5982 | 0.6791 |

| Restriction | Chi-squared | df | p |
|---|---|---|---|
| All three elasticities equal | 87.289 | 2 | <0.0001 |
| Human-capital weight = augmentation weight | 0.002 | 1 | 0.9656 |
| Population elasticity = 1 | 4.795 | 1 | 0.0285 |
| Population elasticity = human-capital weight | 12.966 | 1 | 0.0003 |

The proportionality that ECP newly imposes, between the capability weight and the AI augmentation, is the restriction the data leave intact. The rejected restriction is the common exponent linking scale to quality, which Section 4.6 treats as an accounting question and Section 4.4 reports across the exponent multiverse.

**Table S14. Income saturation of the AI block**

Construct-validation work reports that national AI readiness indices track income closely. This table reports the resulting gradient, the variation that survives inside each income band, and the reranking the AI block produces relative to the HCI+ stock.

| World Bank income group | n | Mean AC | Median AC | Within-group AC p90/p10 | Median absolute rank move from HCI+ stock | Share moving 5 or more |
|---|---|---|---|---|---|---|
| Low income | 16 | 0.095 | 0.096 | 2.98 | 3.5 | 0.375 |
| Lower middle income | 36 | 0.135 | 0.129 | 3.22 | 3.0 | 0.222 |
| Upper middle income | 39 | 0.220 | 0.228 | 2.52 | 1.0 | 0.051 |
| High income | 52 | 0.454 | 0.447 | 2.37 | 3.0 | 0.346 |

The four groups total 143 countries. Ethiopia carries no World Bank income classification in the source panel and is therefore absent from this table alone; it remains in the 144-country sample used for every other result in the manuscript and the supplement.

Across those 143 countries the Spearman correlation between income group and AC is 0.849 with $p = 6.70 \times 10^{-41}$, so the gradient by income is substantial and is reported as measured. Both this correlation and the Kruskal-Wallis test below are computed on the four ranked groups; an economy without an income classification has no position on the scale and is excluded from each rather than entering as a fifth category. Two further features are descriptive: the reranking is not monotone in income, with a Kruskal-Wallis statistic of 22.110 across the four groups and $p = 0.0001$, and variation inside each band remains wide.

**S6. Country coverage**

The final common sample comprises:

ALB (Albania); DZA (Algeria); AGO (Angola); ARG (Argentina); ARM (Armenia); AUS (Australia); AUT (Austria); AZE (Azerbaijan); BHR (Bahrain); BGD (Bangladesh); BLR (Belarus); BEL (Belgium); BEN (Benin); BTN (Bhutan); BIH (Bosnia and Herzegovina); BWA (Botswana); BRA (Brazil); BRN (Brunei Darussalam); BGR (Bulgaria); BFA (Burkina Faso); BDI (Burundi); KHM (Cambodia); CMR (Cameroon); CAN (Canada); CAF (Central African Republic); TCD (Chad); CHL (Chile); CHN (China); COL (Colombia); COD (Congo, Dem. Rep.); COG (Congo, Rep.); CRI (Costa Rica); CIV (Cote d'Ivoire); HRV (Croatia); CYP (Cyprus); CZE (Czechia); DNK (Denmark); DOM (Dominican Republic); ECU (Ecuador); EGY (Egypt, Arab Rep.); SLV (El Salvador); EST (Estonia); SWZ (Eswatini); ETH (Ethiopia); FJI (Fiji); FIN (Finland); FRA (France); GAB (Gabon); GMB (Gambia, The); GEO (Georgia); DEU (Germany); GHA (Ghana); GRC (Greece); GTM (Guatemala); GIN (Guinea); HTI (Haiti); HND (Honduras); HUN (Hungary); ISL (Iceland); IND (India); IDN (Indonesia); IRN (Iran, Islamic Rep.); IRQ (Iraq); IRL (Ireland); ISR (Israel); ITA (Italy); JAM (Jamaica); JPN (Japan); JOR (Jordan); KAZ (Kazakhstan); KEN (Kenya); KOR (Korea, Rep.); KWT (Kuwait); KGZ (Kyrgyz

Republic); LAO (Lao PDR); LVA (Latvia); LBN (Lebanon); LSO (Lesotho); LBR (Liberia); LTU (Lithuania); LUX (Luxembourg); MDG (Madagascar); MWI (Malawi); MYS (Malaysia); MLI (Mali); MLT (Malta); MUS (Mauritius); MEX (Mexico); MDA (Moldova); MNG (Mongolia); MNE (Montenegro); MAR (Morocco); MMR (Myanmar); NAM (Namibia); NPL (Nepal); NLD (Netherlands); NZL (New Zealand); NIC (Nicaragua); NER (Niger); NGA (Nigeria); MKD (North Macedonia); NOR (Norway); OMN (Oman); PAK (Pakistan); PAN (Panama); PRY (Paraguay); PER (Peru); PHL (Philippines); POL (Poland); PRT (Portugal); QAT (Qatar); ROU (Romania); RUS (Russian Federation); RWA (Rwanda); SAU (Saudi Arabia); SEN (Senegal); SRB (Serbia); SYC (Seychelles); SLE (Sierra Leone); SGP (Singapore); SVK (Slovak Republic); SVN (Slovenia); ZAF (South Africa); ESP (Spain); LCA (St. Lucia); VCT (St. Vincent and the Grenadines); SDN (Sudan); SWE (Sweden); CHE (Switzerland); TJK (Tajikistan); TZA (Tanzania); THA (Thailand); TGO (Togo); TUN (Tunisia); TUR (Turkiye); UGA (Uganda); UKR (Ukraine); ARE (United Arab Emirates); GBR (United Kingdom); USA (United States); URY (Uruguay); VNM (Viet Nam); ZMB (Zambia); ZWE (Zimbabwe).

**S7. Falsification and interpretation rules**

The following rules govern interpretation of the measure.

1. A statistically unsupported direct AC interaction cannot be described as established complementarity. The current direct test is null.
2. A country residual cannot be described as causal inefficiency. The external correlates tested here are weak and nonsignificant.
3. Cross-country levels cannot identify the effect of population decline, fertility change, population ageing, or AI adoption over time.

4. HCI+ is prospective; ECP is therefore a scenario-equivalent stock under prevailing conditions.
5. Ranks are planning classifications, not welfare rankings.
6. If independent longitudinal or micro-level outcomes do not support a bottleneck between AI opportunity and conversion capacity, the conjunctive rule must be revised.
7. Because lambda is normative, the benchmark must be accompanied by its components and sensitivity range.

**S8. Reproduction sequence**

The final submission results are regenerated in this order:

1. 00_build_aligned_panel_2024.py
2. 05_build_external_diagnostics.py (the external correlates of Table S10)
3. 06_build_final_measure.py
4. 07_final_analysis.py
5. 08_final_sfa.py
6. 09_make_final_figures.py
7. 11_restriction_and_income_diagnostics.py (Tables S13 and S14)
8. 12_descriptive_and_ledger.py (Table S3.4 and the result ledger)
9. 10_capital_measure_robustness.py (Table S12; requires the Penn World Table 11.0 Stata file, which is redistributed by its owners rather than by this archive)

The one-command runner executes steps 1 through 8 in sequence and stops at the first nonzero exit code. Before it starts it confirms that every input the chain cannot rebuild for itself is present; after it finishes it confirms that every expected result file exists at its full length, that the sample is 144 countries, and that the analysis panel matches its recorded hash. A missing input therefore stops the run rather than producing a shorter table. Step 9 runs separately because it takes the external capital file as an argument, and the archive records that file's own digest so the step can be repeated exactly. A separate read-only program checks every archived file against the manifest, and outputs from the superseded specification are held in a labelled provenance folder outside the result chain.